# Fractal FRW Model within Domain Wall

**D. D. Pawar[1]     D. K. Raut[2]   and W. D. Patil[3]**

[1] **School of Mathematical Sciences, S.R.T.M. University, Nanded-431606, Maharashtra, India**

[2]**Department of Mathematics, Shivaji Mahavidyalya Renapur-413527, Maharashtra, India**

[3]**Department of Mathematics, A. C. Patil college of Engineering, Navi Mumbai-410210, Maharashtra, India**

**E-mail:** dypawar@yahoo.com[1] , dkraut1983@gmail.com[2] , walmikpatil@rediffmail.com[3]

## Abstract

In the present paper an attempt has been made to study the flat fractal Friedmann - Robertson - Walker model filled with domain walls. We have obtained the fractal equation of deceleration parameter and tension of the domain wall. It is observed that, while domain walls exist at early times, they disappear at late time. Finally, some physical parameters of the model are discussed using graphs.

## 1. Introduction

The theory of general relativity (GR) is the most successful theory of gravitation [1], that describes the history of the universe. and it is the basis of the current cosmological models of the universe. The accelerating expansion of the universe has attracted much attention nowadays. The cosmic accelerated expansion of the universe is confirmed using the observational data obtained from the cosmological experiments, supernovae type Ia (SN Ia) [2-6], Large scale structure [7], the Cosmic Microwave Background (CMB) [8-9] etc. and found that the dark energy is the most important factor behind this expansion of the universe. The general theory of relativity suffered many criticisms due to the lack of certain desirable features such as it does not seem to resolve the important problems of dark matter or dark energy in cosmology. So recently some attempts have been made to generalize general relativity. In pursuit of this several modified theories of gravitation have been proposed as an alternative to general relativity such as Brans and Dicke theory of gravitation [10], Barber Self creation theory [11], Saez and Ballester theory of gravitation [12], f(R) gravity [13], f(R,T) gravity[14] etc.

The FRW metric is called as standard model of modern cosmology. It is an exact solution of Einstein's field equation of GR. FRW metric represents isotropic and homogeneous expanding or contracting universe. Many researchers studied the FRW model within modified theories of gravitation. Samantha et al. [15] have studied the bulk viscous fluid in the formalism of f (R, T) theory of gravitation in the frame of flat FRW space-time, Agrawal and Pawar [16] have studied the spatially homogeneous and isotropic FRW model and axially symmetric space-time in f(R) theory of gravity. Sezgin Aygun [17] studied the behavior of cosmological parameter for flat FRW universe with scalar potential in f (R, T) theory of gravity.

In the formation of universe, the domain walls have an important role and it is a topological object. Yilmaz and Atkas [18] have observed that when a discrete symmetry is spontaneously broken in early universe the domain walls might be created. Many researchers are interested in the study of gravitational effects of domain walls. Recently Pawar et al. [19] have investigated thick domain walls in the BD theory of gravitation, Katore and Hatkar [20] have explored Bianchi type -III and Kantowski-sachs domain wall cosmological models in f(R,T) theory of gravitation and found that curved walls are responsible for gravitational collapse of universe, Agrawal and Pawar [21] have studied Bianchi type -V space-time using magnetized domain wall in f(R,T) theory of gravitation and found that in the EM field model is isotropic whereas in the absence it becomes stiff fluid, Pradhan et al. [22] have studied plane symmetric inhomogeneous bulk viscous domain walls in Lyra geometry, Patil et al. [23] have obtained plane symmetric cosmological models of thick domain walls with viscous fluid coupled with electromagnetic field in the framework Lyra geometry.

Fractal cosmology is a term which relates to the appearance of fractal in the study of the universe. There have been numerous cosmological theories such as Causal dynamical triangulations [24], asymptotically safe gravity [25] and loop quantum gravity [26], which suggests that the nature of universe is fractal. Andrei Linde [27] is the first who used fractals in cosmology and suggested eternally existing chaotic inflationary model of universe. The fractal nature of space- time on scale relativity have suggested by Laurent Nottale [28]. Calcagni [29] formulated an effective quantum field theory in a fractal universe with power-counting Renormalizable and which is Lorentz invariant and free from ultraviolet divergence. Karami et al. [30] have studied the fractal flat FRW universe filled with dark energy and dark matter. Mustafa Salti et al. [31] have investigated the extended versions of Holographic and Ricci dark energy models in the fractal theory of gravitation. Dipanjana Das et al. [32] have examined the fractal homogeneous and isotropic FLRW space-time geometry filled with perfect fluid and barotropic equation of state. Lemets and Yerokhin [33] have investigated interacting dark energy models in the frame work of fractal cosmology. The Anisotropic behavior of dark energy in fractal cosmology has studied by Hosseienkhani et al. [34]. Joyce et al. [35] have studied the fractal cosmology in an open universe.

With this motivation, in the present work, an attempt has been made to explain the present (observed) cosmic acceleration in a gravity theory with influence of fractal effects. The paper is organized as follows. In section 2 we discussed fractal gravity formalism. In section 3, we obtained the field equations and derived solutions of field equations. Section 4 describes the brief discussion of the results. Section 5 is devoted to our conclusion.

## 2. Fractal Gravity Formalism

The fractal cosmology is the description of quantum gravity with two features. First the, couplings have the running scaling dimensions, which allows the system to flow from an effective two-dimensional phase in the ultraviolet to an ordinary field theory in the infrared limit. And second, unlike other similar systems, this model maintains Lorentz invariance. The fractal model distribution at small and intermediate scales becomes homogeneous at very large scales and it is fully compatible with the reasonable requirement of all observers and condition of local isotropy around any structure

Assuming matter is minimally coupled with gravity, the total action of Einstein gravity in the fractal space-time [29] can be written as

$$S = S_g + S_m \tag{1}$$

Where, $S_g$ is the gravitational part.

$$S_g = \frac{1}{2K^2}\int d\xi(x)\sqrt{-g}\left(R - 2\wedge - \omega\partial_\mu v\,\partial^\mu v\right) \tag{2}$$

and $S_m$ is the matter part.

$$S_m = \int d\xi(x)\sqrt{-g}\,\mathcal{L}_m \tag{3}$$

Here, $g$ is the determinant of the dimensionless metric $g_{ab}$, $K^2 = 8\pi G$ is Newton's constant, $R$ and $\Lambda$ are the Ricci scalar and cosmological constant respectively. $v$ be a fractional function and $\omega$ is the fractal parameter added because $v$, like the other geometric field $g_{ab}$, is now dynamical. $d\xi(x)$ is the Lebesgue-Stieltjes measure and $\mathcal{L}_m$ is the matter Lagrangian which depends on $g_{ab}$.

Taking the variation of total action (1) with respect to the metric tensor $g_{ab}$, one can obtain the Einstein's field equations in fractal universe as was found in [29]

$$R_{ab} - \frac{1}{2}g_{ab}(R - 2\Lambda) + g_{ab}\frac{\Box v}{v} - \frac{\nabla_a\nabla_b v}{v} + \omega\left(\frac{1}{2}g_{ab}\partial_\sigma v\partial^\sigma v - \partial_a v\,\partial_b v\right) = K^2\,T_{ab} \tag{4}$$

Where $\nabla_a$ is the covariant derivative and $'\Box'$ is usual d' Alembertian operator.

$$\Box\emptyset = \frac{1}{\sqrt{-g}}\,\partial^a\left(\sqrt{-g}\partial_a\emptyset\right) \tag{5}$$

## 3. Fractal FRW Model and its Solutions

In this section we find the exact solution of the FRW space-time in fractal frame work

### 3.1 Fractal FRW space-time

Consider the Friedmann- Robertson –Walker's space-time

$$ds^2 = -dt^2 + A^2(t)\left[\frac{dr^2}{1-kr^2} + r^2\left(d\theta^2 + sin^2\theta\, d\emptyset^2\right)\right] \tag{6}$$

Where the scale factor $A(t)$ is a function of cosmic time t, $k$ is a constant and $k = -1, 0, 1$ represents open, flat and closed FRW universe respectively.

The standard FRW model is in accordance with present day universe and it is referred as standard model of modern cosmology. Recently Naidu et al. [36] have researched FRW viscous fluid cosmological model in f (R, T) gravity. Joachim [37], Katore et al. [38], Boyanorsky et al.

[39], Rosales and Tkach [40] are some of the authors who have investigated several aspects of FRW universe in various contexts.
The corresponding Ricci tensors and Ricci scalar curvature of the FRW model are given by

$$R_{11} = \frac{2k}{A^2}\, g_{11} + \left(3H^2 + \dot{H}\right) g_{11} \tag{7}$$

$$R_{22} = \frac{2k}{A^2}\, g_{22} + \left(3H^2 + \dot{H}\right) g_{22} \tag{8}$$

$$R_{33} = \frac{2k}{A^2}\, g_{33} + \left(3H^2 + \dot{H}\right) g_{33} \tag{9}$$

$$R_{44} = -3\left(H^2 + \dot{H}\right) \tag{10}$$

$$R = 12H^2 + 6\dot{H} + \frac{6k}{A^2} \tag{11}$$

The 44 components of equation (4) is

$$R_{44} - \frac{1}{2} g_{44}(R - 2\Lambda) + g_{44}\frac{\Box v}{v} - \frac{\nabla_4 \nabla_4 v}{v} + \omega\left(\frac{1}{2} g_{44}\partial_\sigma v \partial^\sigma v - \partial_4 v\, \partial_4 v\right) = K^2\, T_{44} \tag{12}$$

Substituting the values of $R_{44}$ , $R$ from equations (10) and (11) in eq. (12), we get

$$H^2 + H\frac{\dot{v}}{v} - \frac{1}{6}\omega\dot{v}^2 + \frac{k}{A^2} = \frac{K^2}{3}\, T_{44} + \frac{\Lambda}{3} \tag{13}$$

Taking the trace of eq. (4) gives

$$-R = 4\,\Lambda + 3\frac{\Box v}{v} + \omega\partial_a v \partial^a v = K^2\, T_a^a \tag{14}$$

With the help of equations (11) and (13), equation (14) reduces to

$$2H^2 + \dot{H} - \frac{1}{2}\frac{\Box v}{v} + \frac{1}{6}\omega\dot{v}^2 + \frac{k}{A^2} = \frac{-K^2}{6}\, \mathrm{T}_a^a + \frac{2\Lambda}{3} \tag{15}$$

Where overhead dot $(\cdot)$ denotes derivative with respect to time t and $H = \frac{\dot{A}}{A}$ be the generalized mean Hubble parameter. Equation (13) and (15) are the Friedmann equations in the fractal universe.

The energy momentum tensor of domain wall is

$$T_{ab} = \rho(g_{ab} + w_a\, w_b) + p\, g_{ab} \tag{16}$$

together with

$$w^a\, w^b = -1 \tag{17}$$

Where p is pressure normal to the plane of domain wall, $\rho$ is the energy density of the domain wall and $w^a$ is four velocities vector in the same direction. The non-vanishing components of $T_{ab}$ can be obtained using co-moving co-ordinate system as

$$T_1^1 = T_2^2 = T_3^3 = \rho\,, T_4^4 = p \tag{18}$$

Domain walls are the topological defects associated with spontaneous symmetry breaking of the universe. Domain walls are cosmologically important. The study of the domain walls has gained renewed cosmological interest due to their application in structure formation in the universe [41]-[42], Rahaman et. al [43], Adhav et al. [44], Sahoo et al. [45], Shaikh et al. [46], Wang [47] are some of the authors who researched several aspects of domain walls in various circumstances. Reddy and Naidu [48] have shown that the domain walls do not survive in scalar co-variant theory.

The equations (13) and (15), for the metric (6) with the help equation of eq. (18) reduces to

$$H^2 + H\frac{\dot{v}}{v} - \frac{1}{6}\omega\dot{v}^2 + \frac{k}{A^2} = \frac{-K^2}{3}\,p + \frac{\Lambda}{3} \tag{19}$$

$$H^2 + \dot{H} - \mathrm{H}\frac{\dot{v}}{v} - \frac{1}{2}\frac{\Box v}{v} + \frac{1}{3}\omega\dot{v}^2 = \frac{K^2}{6}\,(\mathrm{p} - 3\rho) + \frac{\Lambda}{3} \tag{20}$$

Note that by taking $v = 1$ in eq. (19) and (20), we get the standard Friedmann equations in Einstein's general theory of relativity.

$$H^2 = \frac{-K^2}{3}\,p + \frac{\Lambda}{3} - \frac{k}{A^2} \tag{21}$$

$$H^2 + \dot{H} = \frac{K^2}{6}\,(p - 3\rho) + \frac{\Lambda}{3} \tag{22}$$

In a fractal space- time the purely gravitational constraint is given by the following equation

$$3H^2 + H^{\cdot} + H\frac{\dot{v}}{v} + \frac{\Box v}{v} + \frac{2k}{A^2} - \omega(v\Box v - \dot{v}^2) = 0 \tag{23}$$

For flat FRW metric $(k = 0)$, taking monomial form of the fractional function $v = t^{-\gamma}$ [29], where $\gamma = 4(1-\alpha)$ is the fractal dimension of $\xi$ and $\alpha$ is positive parameter roughly corresponds to the fraction of states preserved at a given time during the evolution of the system, one can obtain.

$$\frac{\Box v}{v} = \frac{\gamma}{t}\left(3H - \frac{1+\gamma}{t}\right) \tag{24}$$

The monomial form of the fractional function $v$ is consistent with recent observations particularly for describing the early and intermediate era of evolution. And if $v$ = 1, it turns out that predictions of the theory approaches to those of standard Einstein's theory.

For simplicity, we choose $\Lambda = 0$ (no cosmological constant) and units are chosen such that $8\pi G = 1$ , in the UV regime (Small scale structure $\gamma = 2$) ,equations (21), (22) and (23) can be written as

$$H^2 - \frac{2}{t}H - \frac{2\omega}{3t^6} = \frac{-1}{3}\,p \tag{25}$$

$$H^2 + \dot{H} - \frac{1}{t}H + \frac{3}{t^2} + \frac{4\omega}{3t^6} = \frac{1}{6}\left(p - 3\rho\right) \tag{26}$$

$$3H^2 + \dot{H} + \left(2 + \frac{3\omega}{t^4}\right)\frac{2}{t}H = \frac{6}{t^2} + \frac{10\omega}{t^6} \tag{27}$$

We define some physical parameters of the model

The average scale factor $a$ is defined as

$$a = \sqrt[3]{A^3} = A \tag{28}$$

The volume scale factor is given as

$$V = a^3 = A^3 \tag{29}$$

The scalar expansion $\theta$ is given by

$$\theta = 3H = H_1 + H_2 + H_3 = 3\frac{\dot{A}}{A} \tag{30}$$

Where $H_1 = H_2 = H_3 = \frac{\dot{A}}{A}$ are the directional Hubble parameters in the directions of $x$ , $y$ and $z$ - axes respectively.

The measure of cosmic acceleration of the universe expansion is a deceleration parameter. which is defined as

$$q = \frac{-A\ddot{A}}{\dot{A}^2} \tag{31}$$

## 3.2 Solutions of the Field Equations

If $\omega = 0$ , then from eq. (27) the scale factor $A(t)$ is obtained as

$$A(t) = \frac{(t^9 - c)^{\frac{1}{3}}}{t^2} \tag{32}$$

Where, $c$ is the constant of integration.

The Hubble parameter is given by

$$H = \frac{(t^9 + 2c)}{t(t^9 - c)} \tag{33}$$

Inserting the value of $A$ from eq. (32) into the equations (29) and (30), we get the required volume of the universe and expansion scalar.

$$V = \frac{(t^9 - c)}{t^6} \tag{34}$$

$$\theta = \frac{3(t^9 + 2c)}{t(t^9 - c)} \tag{35}$$

Solving eq. (31), we can obtain fractal deceleration parameter

$$q = \left(\frac{-2}{3}\right)\frac{(t^9 - c)(t^9 - 7c)}{(t^9 + 2c)^2} \tag{36}$$

Solving eq. (21) and (22), we get

$$p = 3\,\frac{(t^9+2c)(t^9-4c)}{t^2(t^9-c)^2} \tag{37}$$

$$\rho = -3\,\frac{\left(t^9+\left(10+3\sqrt{10}\right)c\right)\left(t^9-\left(10-3\sqrt{10}\right)c\right)}{t^2(t^9-c)^2} \tag{38}$$

Assuming $\omega \neq 0$ in eq. (27) and solving it, gives the scale factor $A$ [29] as

$$A = \frac{1}{t^2}\,M\left(\frac{11}{4};\frac{13}{4};\frac{3\omega}{2t^4}\right)^{\frac{1}{3}} \tag{39}$$

Where M is Kummer's confluent hypergeometric function of the first kind. Which is given by

$$M(a;b;z) = \frac{\Gamma(b)}{\Gamma(a)}\sum_{m=0}^{\infty}\frac{\Gamma(a+m)}{\Gamma(b+m)}\,\frac{z^m}{m!} \tag{40}$$

Using the value of $A$ , the Hubble parameter is given by

$$H = \frac{-2}{t} + \frac{4\omega}{13t^5}\,\frac{M\left(\frac{11}{4};\frac{17}{4};\frac{3\omega}{2t^4}\right)}{M\left(\frac{11}{4};\frac{13}{4};\frac{3\omega}{2t^4}\right)} - \frac{2\omega}{t^5} \tag{41}$$

Substituting the value of $A$ from eq. (39) into the equations (29) and (30), volume of the universe and expansion scalar are found to be

$$V = \frac{1}{t^6}M\left(\frac{11}{4};\frac{13}{4};\frac{3\omega}{2t^4}\right) \tag{42}$$

$$\theta = \frac{-6}{t} + \frac{12\omega}{13t^5}\,\frac{M\left(\frac{11}{4};\frac{17}{4};\frac{3\omega}{2t^4}\right)}{M\left(\frac{11}{4};\frac{13}{4};\frac{3\omega}{2t^4}\right)} - \frac{6\omega}{t^5} \tag{43}$$

solving eq. (31), deceleration parameter $q$ in fractal space is given by

$$q = \frac{-N}{D} \tag{44}$$

Where

$$N = 169(3t^4+4\omega)(t^4+\omega)M\left(\frac{11}{4};\frac{13}{4};\frac{3\omega}{2t^4}\right)^2 + 52\omega(2t^4+\omega)M\left(\frac{11}{4};\frac{17}{4};\frac{3\omega}{2t^4}\right)M\left(\frac{11}{4};\frac{13}{4};\frac{3\omega}{2t^4}\right) - 16\omega^2 M\left(\frac{11}{4};\frac{17}{4};\frac{3\omega}{2t^4}\right)^2 \tag{45}$$

$$D = 338(t^4+\omega)^2M\left(\frac{11}{4};\frac{13}{4};\frac{3\omega}{2t^4}\right)^2 - 104\omega(t^4+\omega)\,M\left(\frac{11}{4};\frac{17}{4};\frac{3\omega}{2t^4}\right)M\left(\frac{11}{4};\frac{13}{4};\frac{3\omega}{2t^4}\right) - 8\omega^2M\left(\frac{11}{4};\frac{17}{4};\frac{3\omega}{2t^4}\right)^2 \tag{46}$$

Now solving equations (25) and (26) for the $p$ and $\rho$ one obtains

$$p = \frac{-2(3t^4+2\omega)(4t^4+3\omega)}{t^{10}} - \frac{48\omega^2}{13^2 t^{10}} \frac{M\left(\frac{11}{4};\frac{17}{4};\frac{3\omega}{2t^4}\right)^2}{M\left(\frac{11}{4};\frac{13}{4};\frac{3\omega}{2t^4}\right)^2} + \frac{24(t^4+2\omega)}{13t^{10}} \frac{M\left(\frac{11}{4};\frac{17}{4};\frac{3\omega}{2t^4}\right)}{M\left(\frac{11}{4};\frac{13}{4};\frac{3\omega}{2t^4}\right)} \tag{47}$$

$$\rho = \frac{-2(5t^4+6\omega)(3t^4+\omega)}{t^{10}} + \frac{48\omega^2}{13^2 t^{10}} \frac{M\left(\frac{11}{4};\frac{17}{4};\frac{3\omega}{2t^4}\right)^2}{M\left(\frac{11}{4};\frac{13}{4};\frac{3\omega}{2t^4}\right)^2} \tag{48}$$

The tension of the domain wall is given by

$$\sigma_d = \frac{2}{t^{10}} \left[ \begin{array}{c} (5t^4 + 6\omega)(3t^4 + \omega) + \frac{48\omega^2}{13} \frac{M\left(\frac{11}{4};\frac{17}{4};\frac{3\omega}{2t^4}\right)^2}{M\left(\frac{11}{4};\frac{13}{4};\frac{3\omega}{2t^4}\right)^2} - \frac{1}{\gamma}(3t^8 + 12t^4 + 4\omega^2) \\ +12\omega(3t^4 + 2\omega) \frac{M\left(\frac{11}{4};\frac{17}{4};\frac{3\omega}{2t^4}\right)^2}{M\left(\frac{11}{4};\frac{13}{4};\frac{3\omega}{2t^4}\right)^2} \end{array} \right] \tag{49}$$

At early times and late time, we distinguish the result for positive and negative values of fractal parameter $\omega$ and using asymptotic form of Kummer's confluent hypergeometric function

When $z \to -\infty$

$$M(a;b;z) \sim \frac{\Gamma(b)}{\Gamma(b-a)} (-z)^{-a} + \sum_{m=1}^{N} \frac{\Gamma(a+m)}{\Gamma(a)} \frac{\Gamma(b)}{\Gamma(b-a-m)} \frac{z^{-m}}{m!} \tag{50}$$

When $z \to \infty$

$$M(a;b;z) \sim \frac{\Gamma(b)}{\Gamma(a)} e^z z^{a-b} + e^z z^{a-b} \sum_{m=1}^{N} \frac{\Gamma(b-a+m)}{\Gamma(b-a)} \frac{\Gamma(b)}{\Gamma(a-m)} \frac{(-z)^{-m}}{m!} \tag{51}$$

we have following cases.

**Case I:** when $t \to 0$ (at early time) and $\omega > 0$

Using equation (51) in equations (41) - (49) we obtain

$$V \sim constant \times \frac{e^{\frac{3\omega}{2t^4}}}{t^4} \tag{52}$$

$$H \sim \frac{-2\omega}{t^5} \tag{53}$$

$$\theta \sim \frac{-6\omega}{t^5} \tag{54}$$

$$p \sim \frac{-12\omega^2}{t^{10}} \tag{55}$$

$$\rho \sim \frac{-12\omega^2}{t^{10}} \tag{56}$$

$$q \sim -1 - \frac{5t^4}{2\omega} \tag{57}$$

$$\sigma_d \sim \frac{-8\omega^2}{\gamma\, t^{10}}(4\gamma - 1) \tag{58}$$

**Case II:** when $t \to 0$ (at early time) and $\omega < 0$

Using equation (50) in equations (41) - (49) we obtain

$$V \sim constant \times t^5 \tag{59}$$

$$H \sim \frac{5}{3t} \tag{60}$$

$$\theta \sim \frac{5}{t} \tag{61}$$

$$p \sim \frac{12\omega}{t^6} \tag{62}$$

$$\rho \sim \frac{-46\omega}{t^6} \tag{63}$$

$$q \sim \frac{-2}{5} \tag{64}$$

$$\sigma_d \sim \frac{-32\omega^2}{\gamma\, t^{10}} \tag{65}$$

**Case-III:** At late time $t \to \infty$ and for $\omega > 0$

$$V \sim constant \times \frac{1}{t^6} \tag{66}$$

$$H \sim \frac{-2}{t} \tag{67}$$

$$\theta \sim \frac{-6}{t} \tag{68}$$

$$p \sim \frac{-24}{t^2} \tag{69}$$

$$\rho \sim \frac{-30}{t^2} \tag{70}$$

$$q \sim \frac{-3}{2} \tag{71}$$

$$\sigma_d \sim \frac{-6}{\gamma\, t^2}(5\gamma + 1) \tag{72}$$

For late time $t \to \infty$ and for $\omega < 0$ results are similar to Case-III.

# 4. Discussion

In this section some important physical parameters of the solutions such as Volume $V$, Hubble parameter $H$ , Expansion scalar $\theta$, deceleration parameter $q$ are evaluated for the models and presented in terms of graphs. The graphs are drawn for particular values of fractal parameter.

For $\omega = 0$ , when $c > 0$ from eq. (29) and eq. (31), it will be observed that at $t = c^{\frac{1}{9}}$ volume is zero, when $t > c^{\frac{1}{9}}$ volume is an increasing function of cosmic time, which results into expanding universe. The deceleration parameter is negative from $t = c^{\frac{1}{9}}$ to $t_1 = (7c)^{\frac{1}{9}}$ and it is positive for $t > t_1$, Hence universe is expanding and accelerating from $t = c^{\frac{1}{9}}$ to $t_1 = (7c)^{\frac{1}{9}}$ and for $t > t_1$, universe is expanding and decelerating. When $c = 0$, from eq. (29) and eq. (31), it is observed that volume is a positive valued decreasing function of cosmic time and deceleration parameter is negative throughout the evolution of the universe, which shows universe is contracting and accelerating.

### 4.1 When $t \to 0$ (at early time) and $\omega > 0$

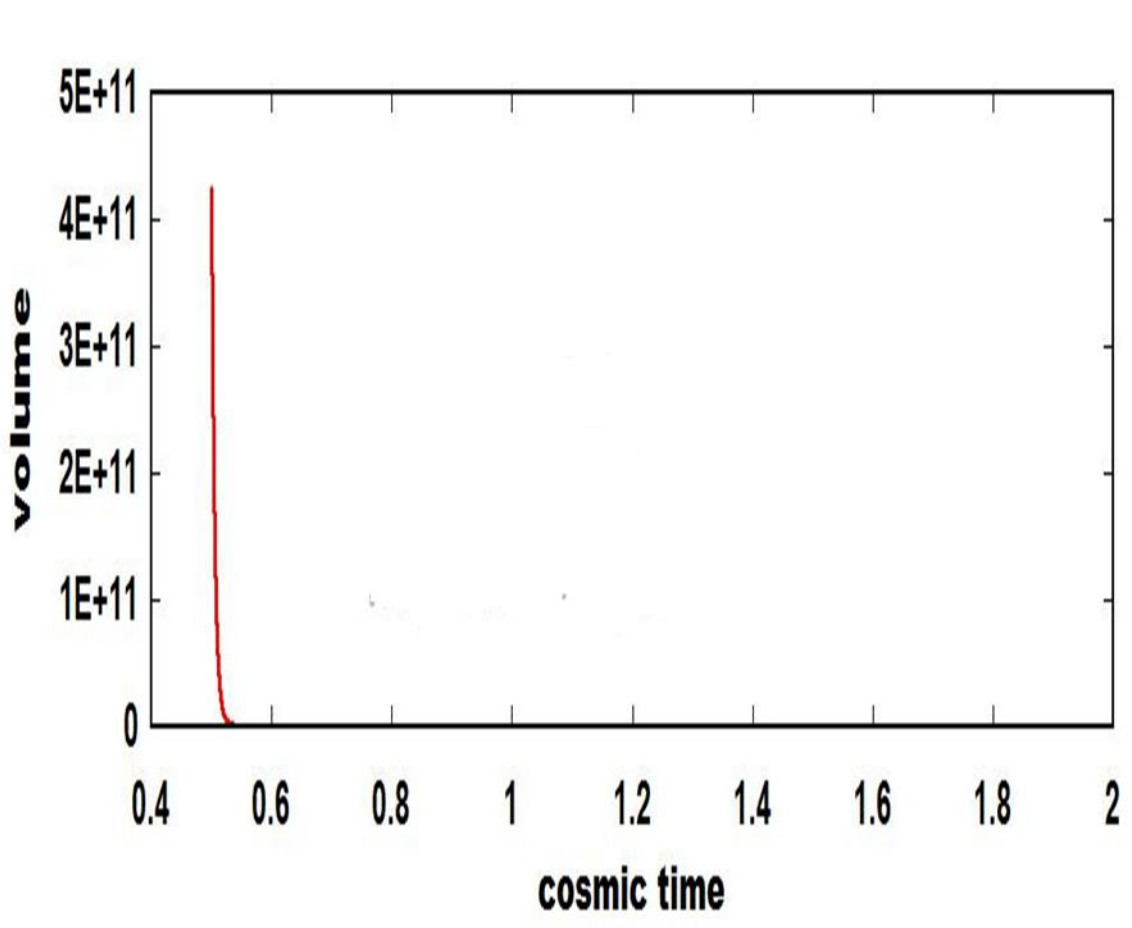

Fig. 1 (a) Variation of volume vs. cosmic time for $\omega = 1$.

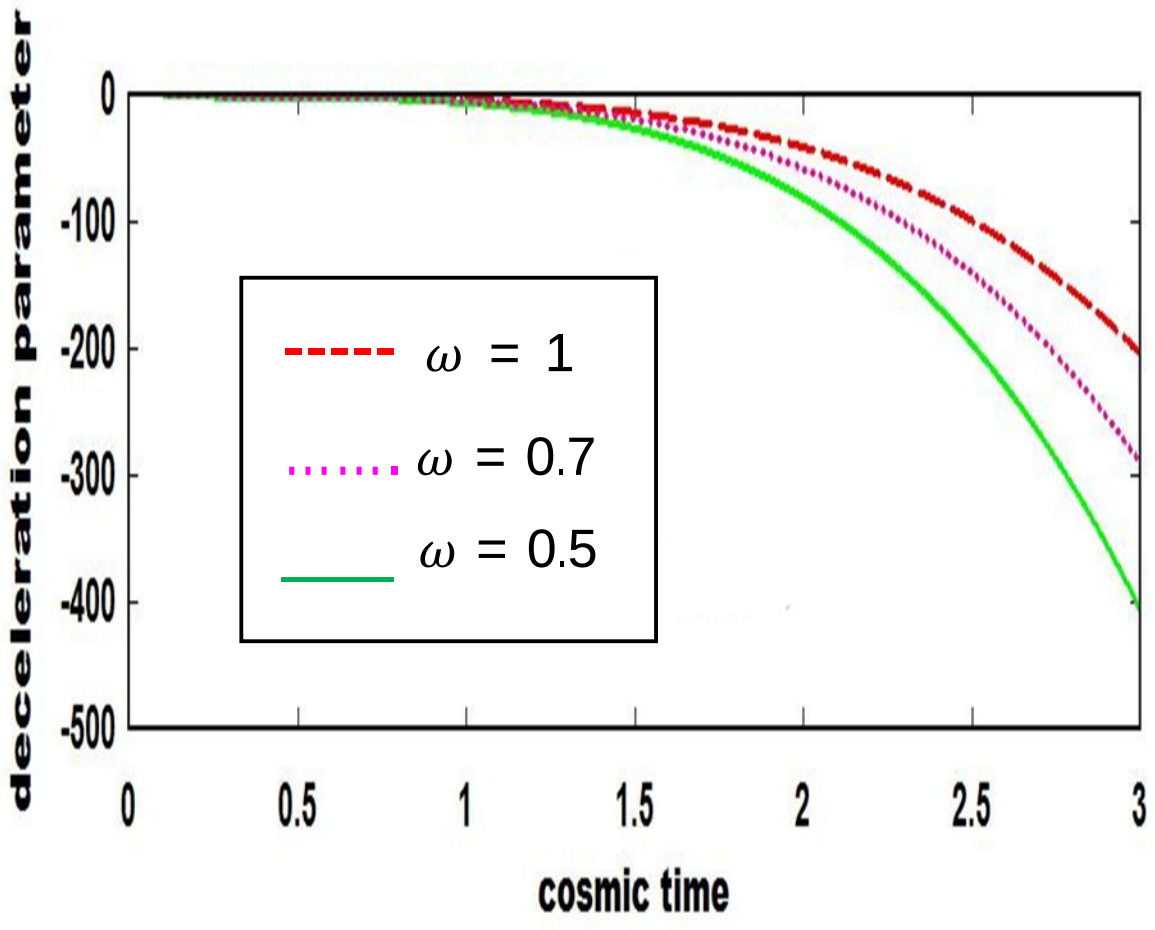

Fig. 1 (b) Variation of deceleration parameter vs. cosmic time for $\omega = 1$

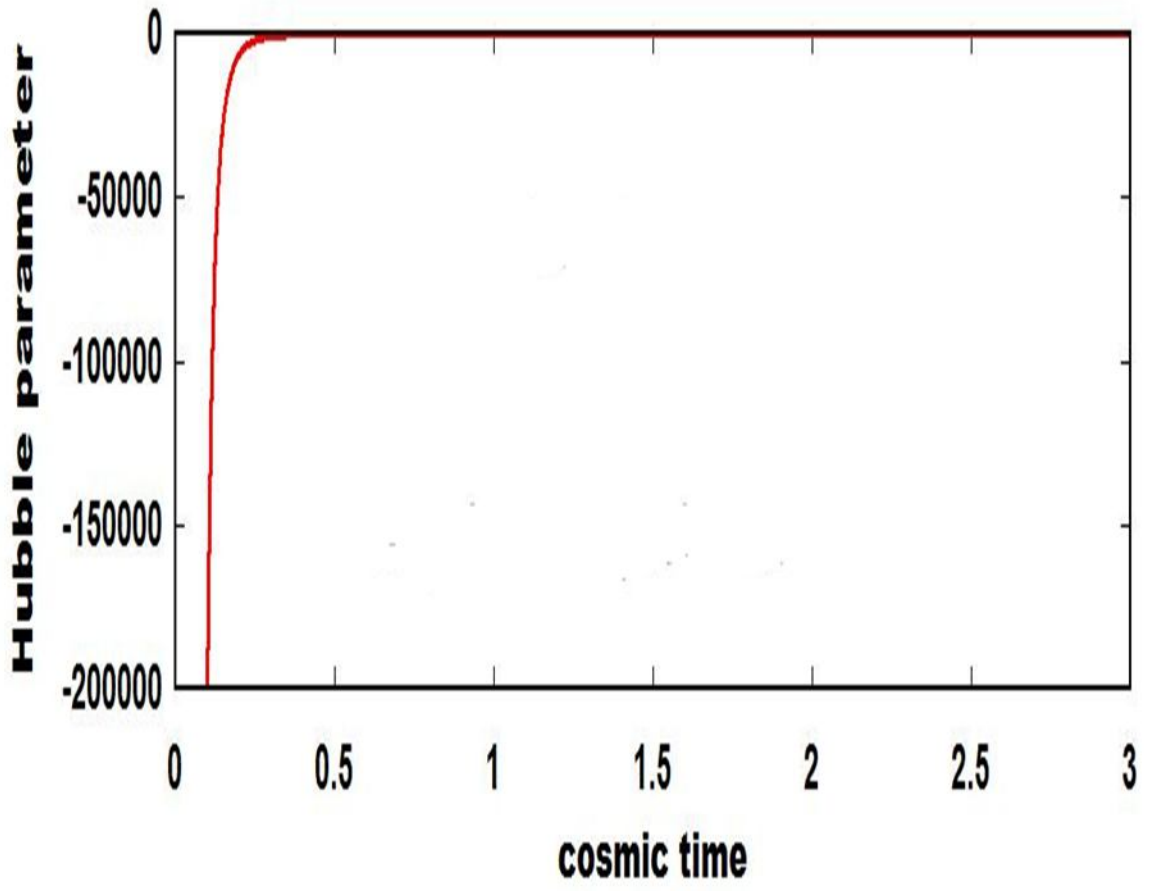

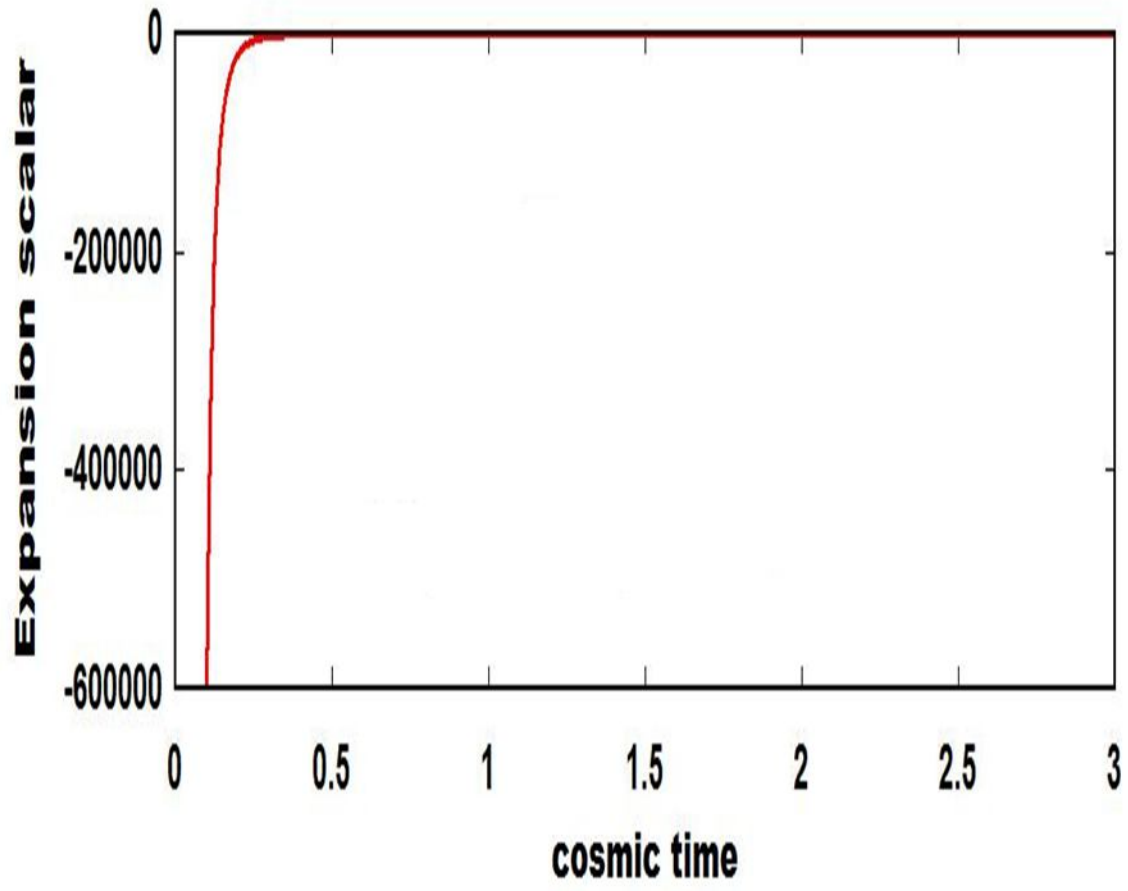

**Fig. 1 (c) Variation of Hubble parameter vs. cosmic time for $\omega = 1$ .** **Fig. 1(d) Variation of expansion scalar vs. cosmic time for $\omega = 1$.**

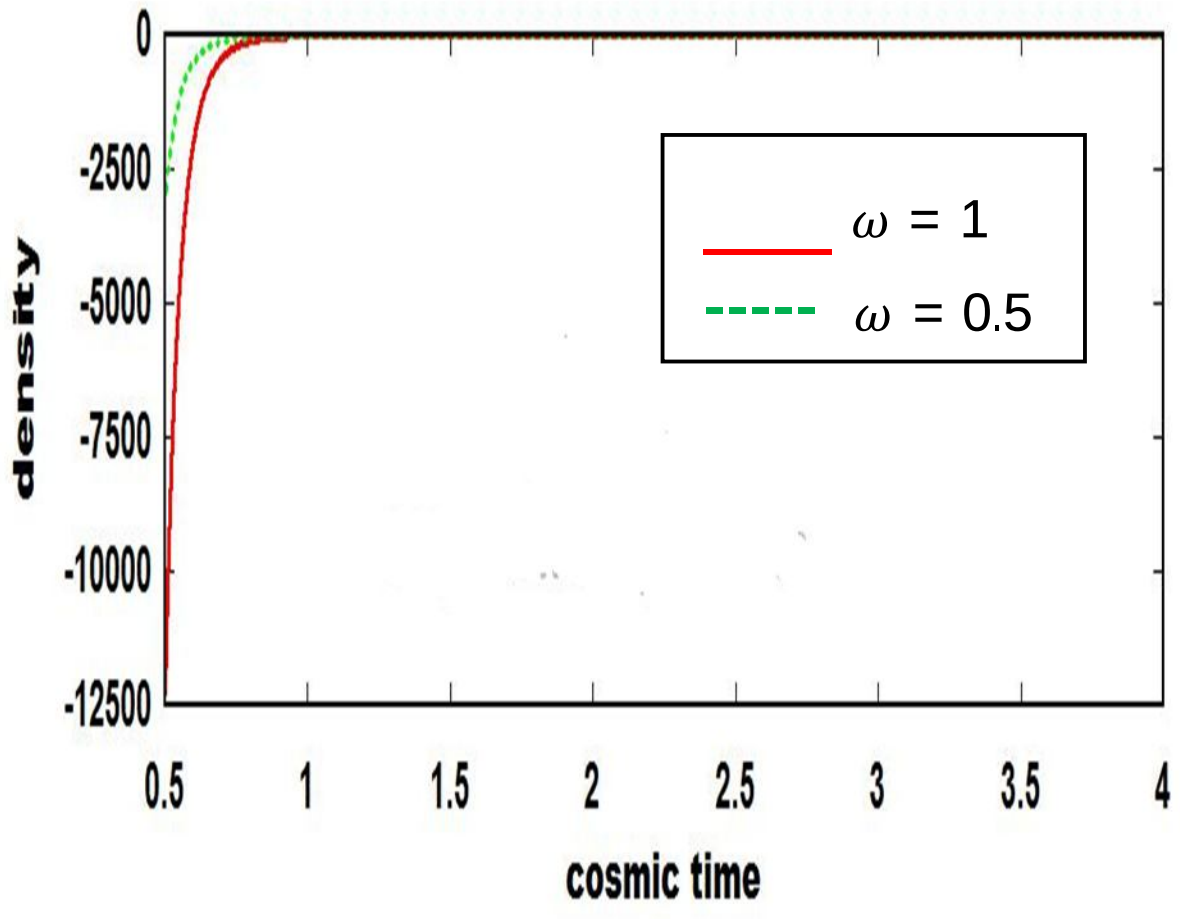

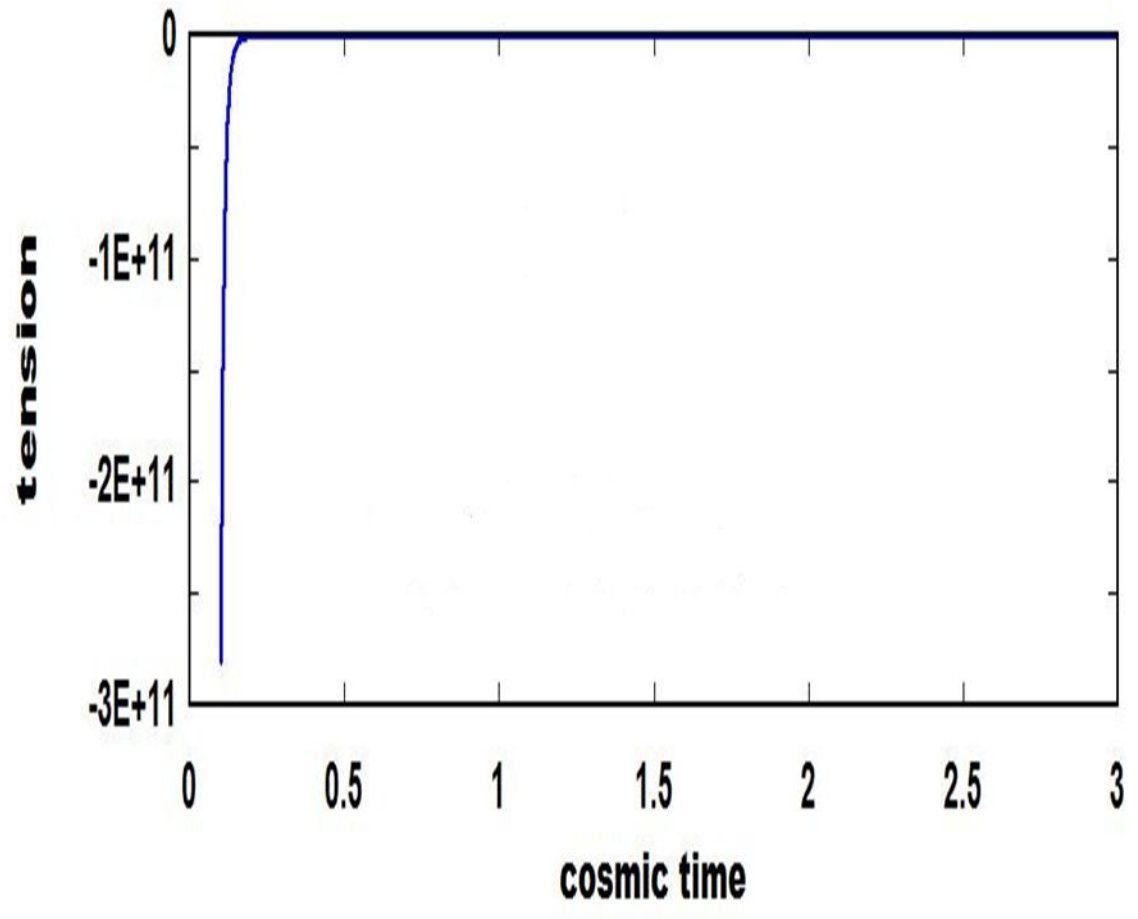

**Fig. 1 (e) Variation of energy density vs. cosmic time for $\omega = 1$.** **Fig. 1(f) Variation of tension vs. cosmic time for $\omega = 1$.**

From Fig.1 (a) it is observed that the volume $V$ of our model is decreases with time $t$ indicating that universe is contracting. The deceleration parameter $q$ is negative throughout evolution of the universe which results into accelerating universe. At $t = 0$ , $q = -1$ as shown in the plot of Fig. 1(b). From eq. (47) and eq. (48), it is observed that the pressure $p$ and energy density $\rho$ both are equal, which represents self – graviting or stiff domain walls. Fig. 1(e) is the plot of energy density versus cosmic time for $\omega = 0.5, \omega = 1$. Initially when time is zero then energy density and pressure both approach to negative infinity. For this case we received the initial singularity. Both density and pressure increase with time and approach to zero for large values of time. From Fig. 1(c) and Fig. 1(d), it is observed that initially at $t = 0$ , the Hubble parameter $H$ and expansion scalar $\theta$ are approaches to negative infinity and vanishes for large time. Fig. 1(f) represents the variation of tension of domain walls versus cosmic time. The tension of the domain walls is negative valued function of cosmic time and vanishes for large time. The null energy condition is violated for this model contrary to the standard model of general relativity.

## 4.2 When $t \to 0$ (at early time) and $\omega < 0$

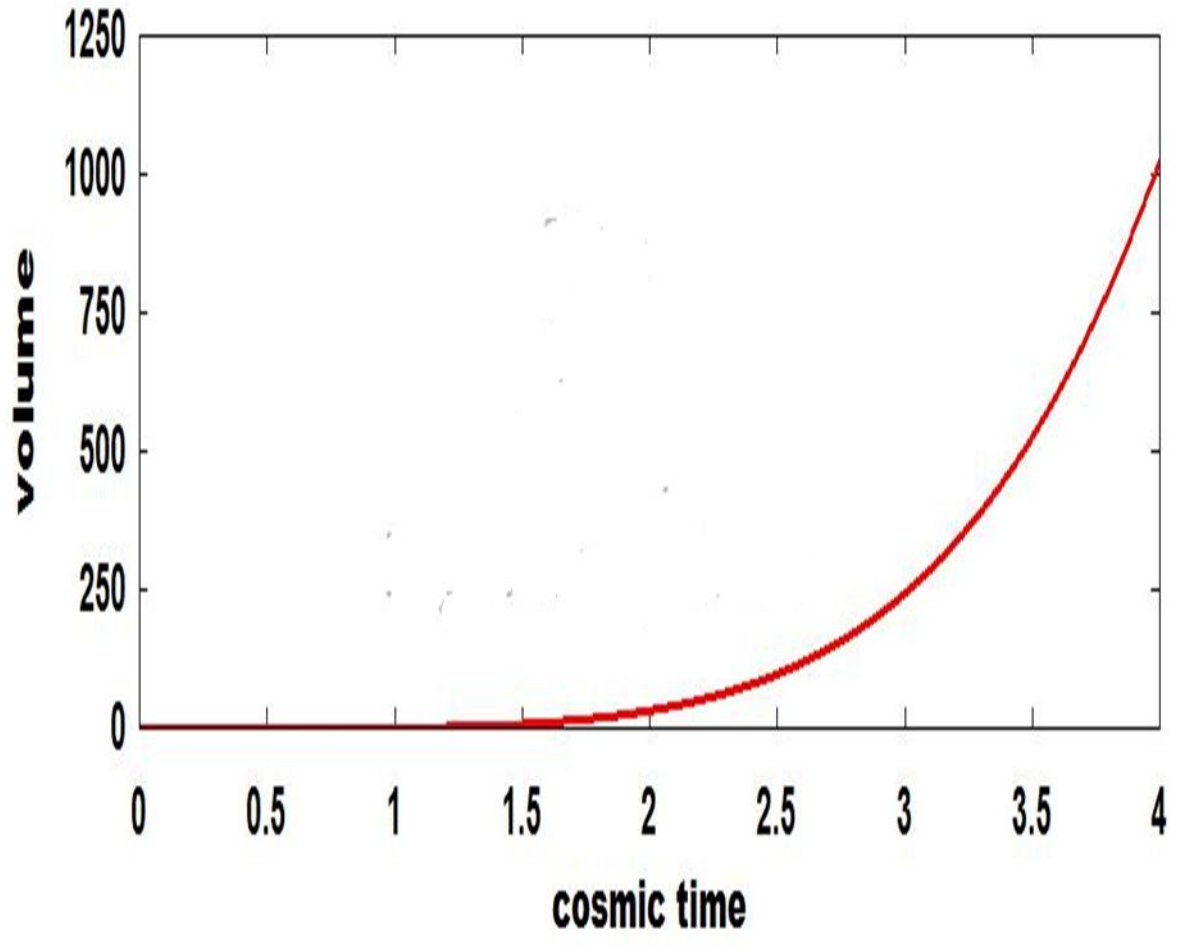

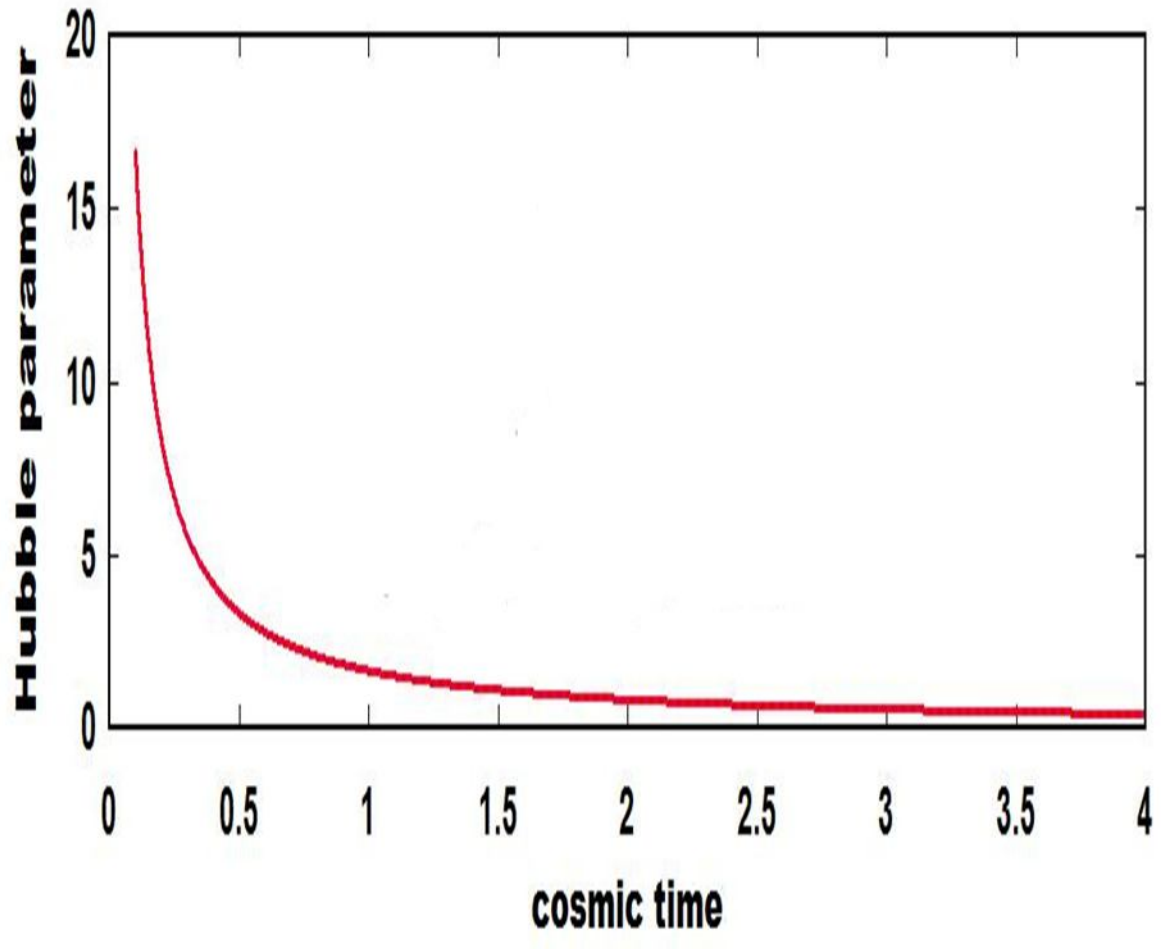

**Fig. 2 (a) Variation of volume vs. cosmic time for $\omega = -1$.** **Fig. 2 (b) Variation of Hubble parameter vs. cosmic time for $\omega = -1$.**

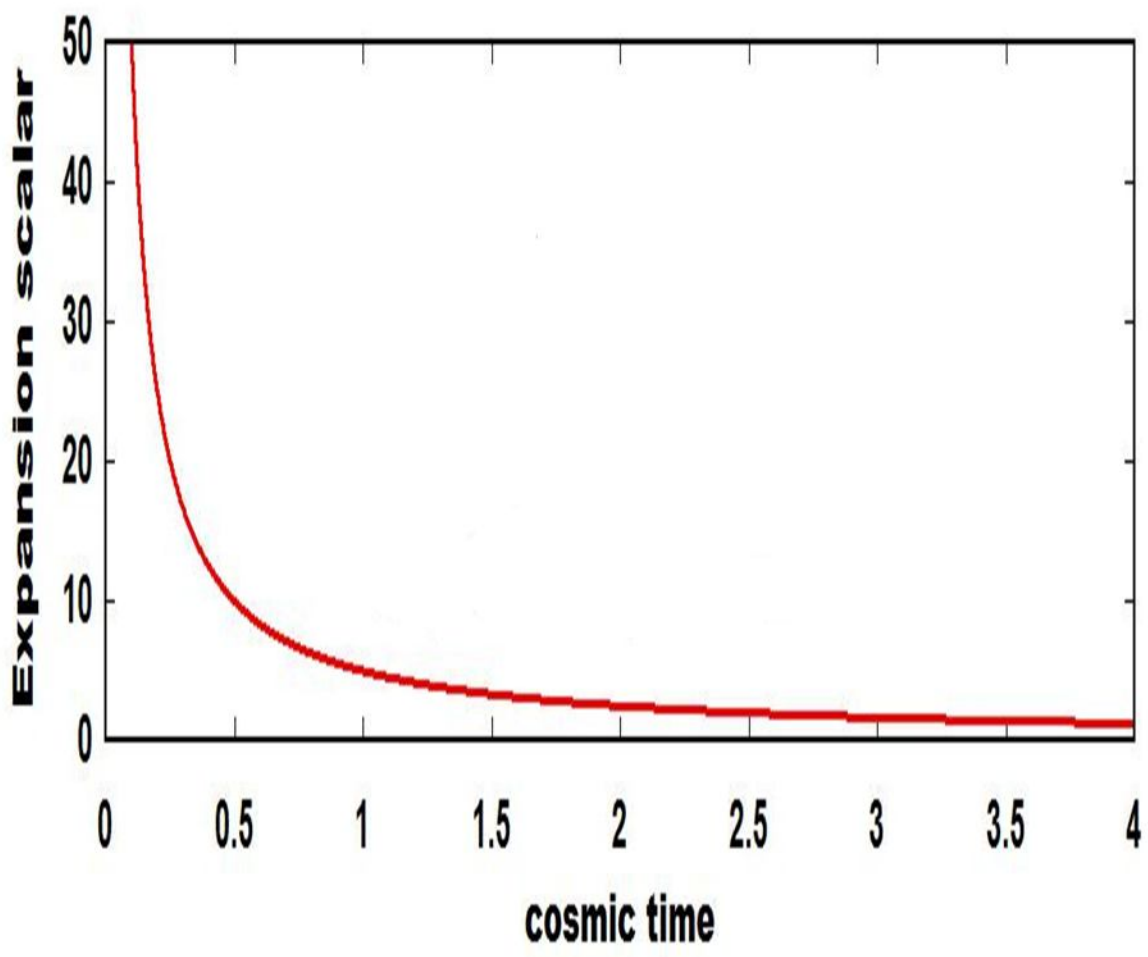

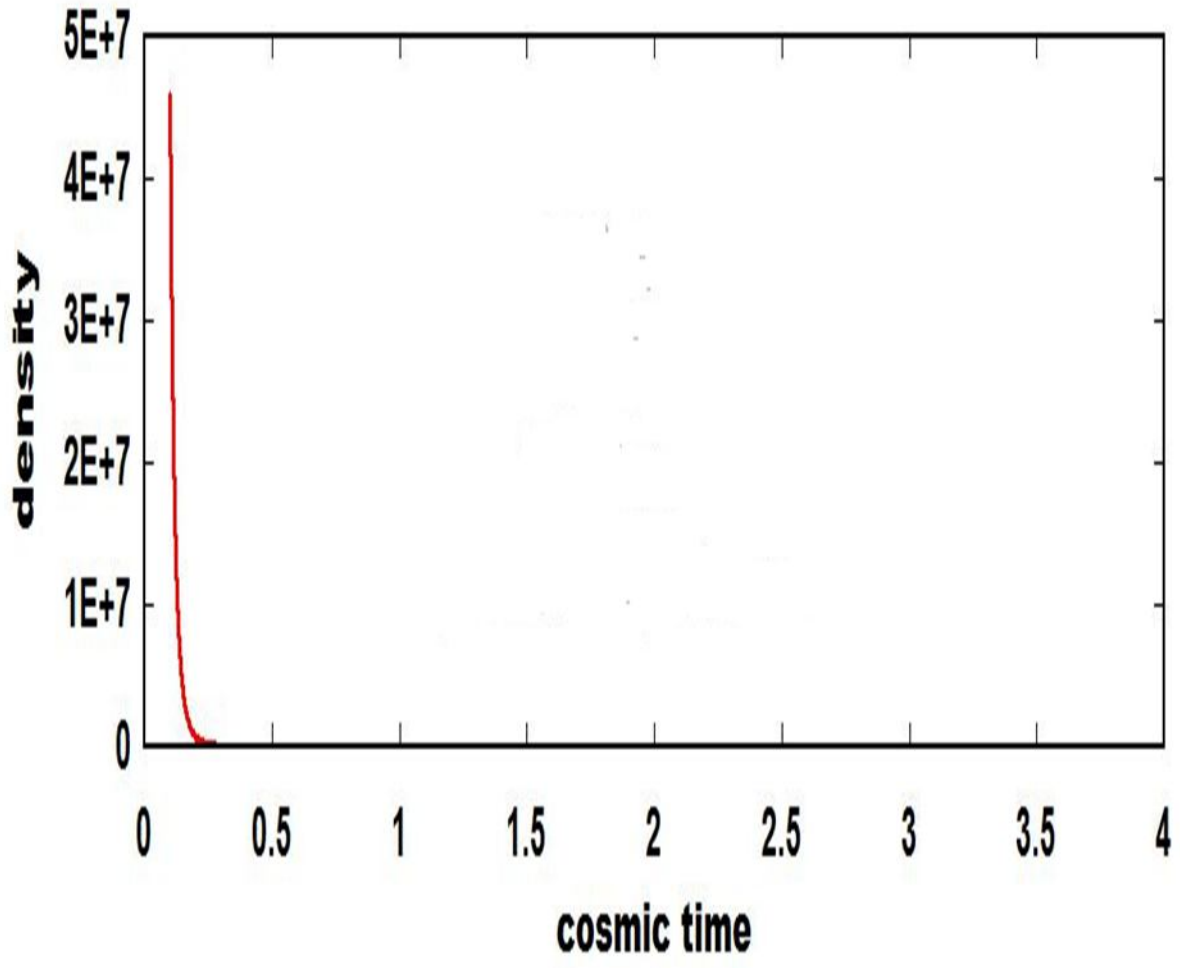

**Fig. 2(c) Variation of expansion scalar vs. cosmic time.** **Fig .2 (d) Variation of density vs. cosmic time.**

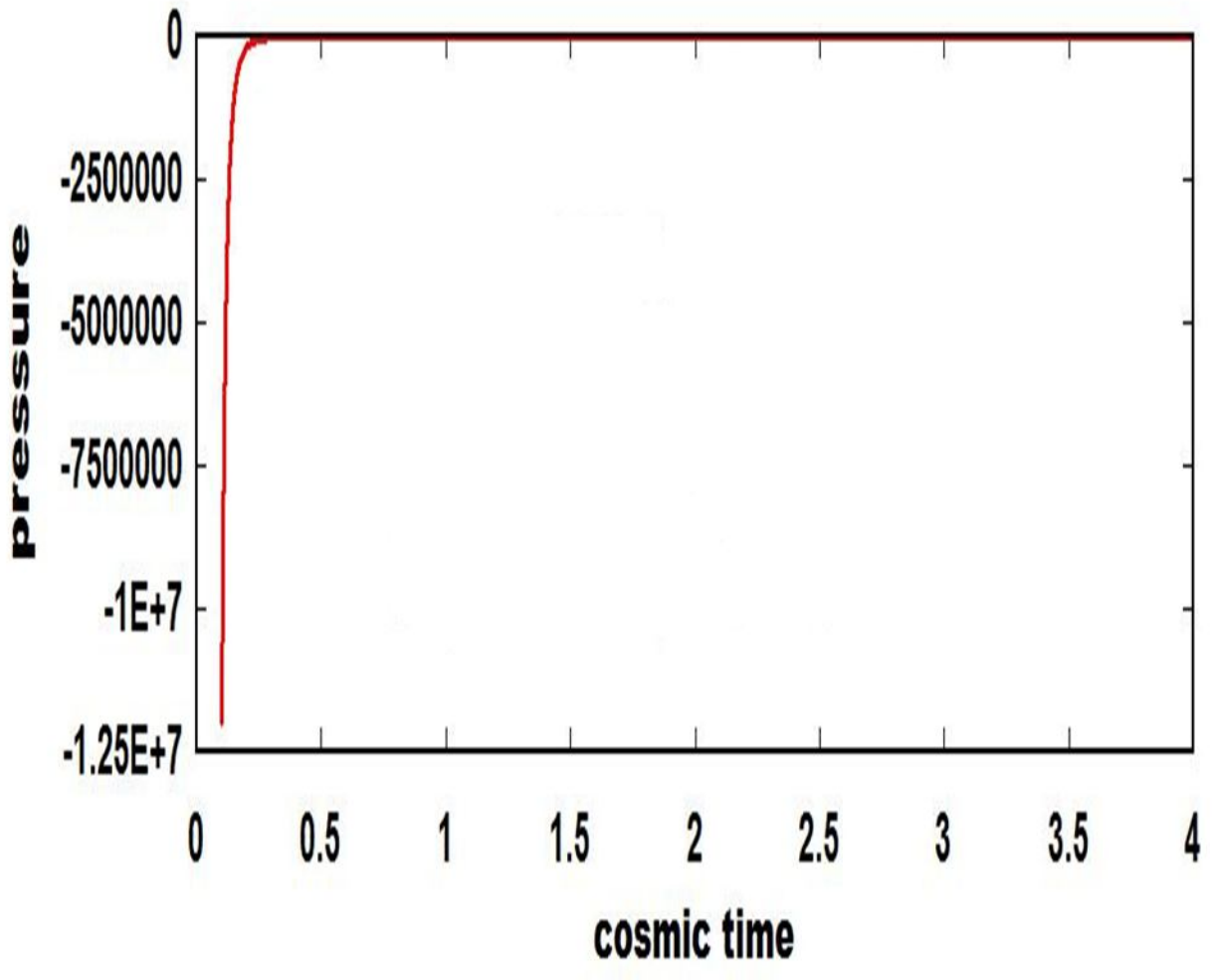

**Fig. 2 (e) Variation of pressure vs. cosmic time for $\omega = -1$.**

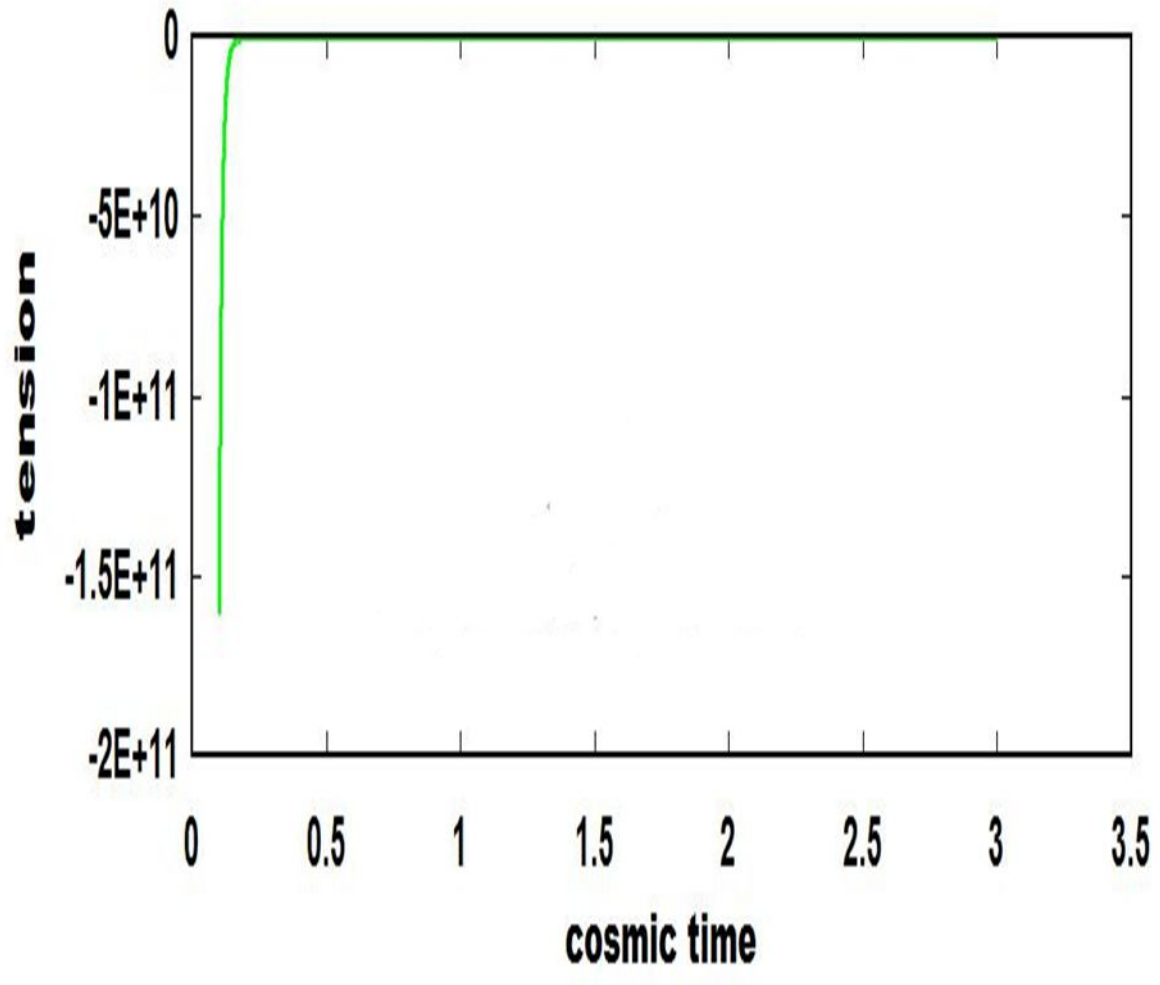

**Fig. 2 (f) Variation of tension vs. cosmic time for $\omega = -1$.**

Figure 2 (a) depicts the variation of spatial volume versus cosmic time. It is observed that at the initial moment i.e. when time $t = 0$, the spatial volume of the model is zero and increases with increase in time, hence the present model is expanding. At $t = 0$ , the expansion of the model starts with Big Bang singularity. From eq. (64), the deceleration parameter for the present model is negative and constant throughout the evolution of the universe, indicating the accelerating expansion of the universe which is well matched with present universe. The Hubble parameter and expansion scalar are extremely large near the origin. Both decrease with increase in time as shown in Fig. 2(b) and Fig. 2(c). it is observed that the rate of expansion is faster at the beginning and slows down in later stage. Fig. 2(d) and Fig. 2(e) are the plots of energy density and cosmic pressure verses cosmic time. It is observed that the energy density of the model is infinite at the origin of universe and becomes zero as $t \to \infty$ . The cosmic pressure is negative-valued increasing function of time. Fig. 2(f) shows that the domain walls are invisible in this model. This is instance with the argument of Zeldovich [49]. It is observed that the Hubble parameter, expansion scalar, cosmic pressure, energy density of the model is zero for large time, hence for the large time present model gives the empty space.

### 4.3 When $t \rightarrow \infty$ (at late time) and $\omega > 0$

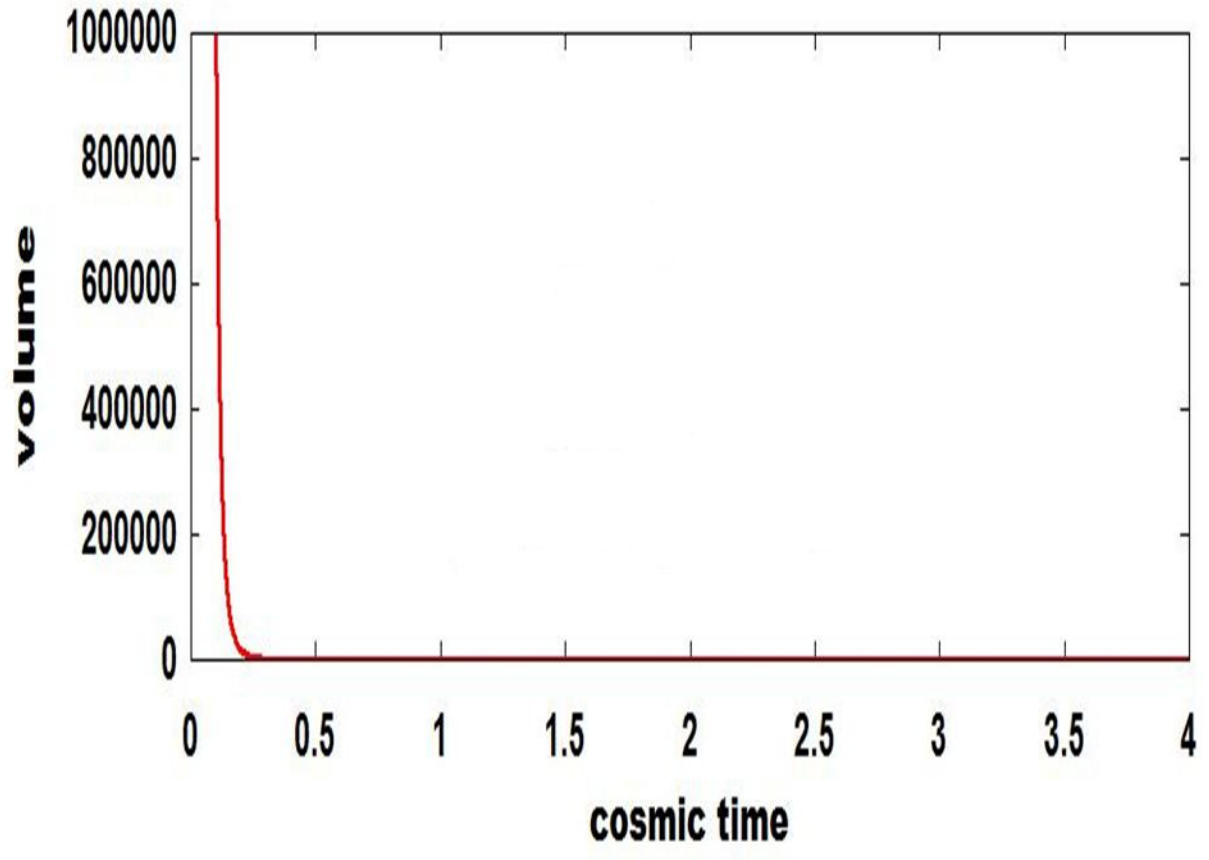

**Fig. 3 (a) Variation of volume vs. cosmic time for $\omega = 1$.**

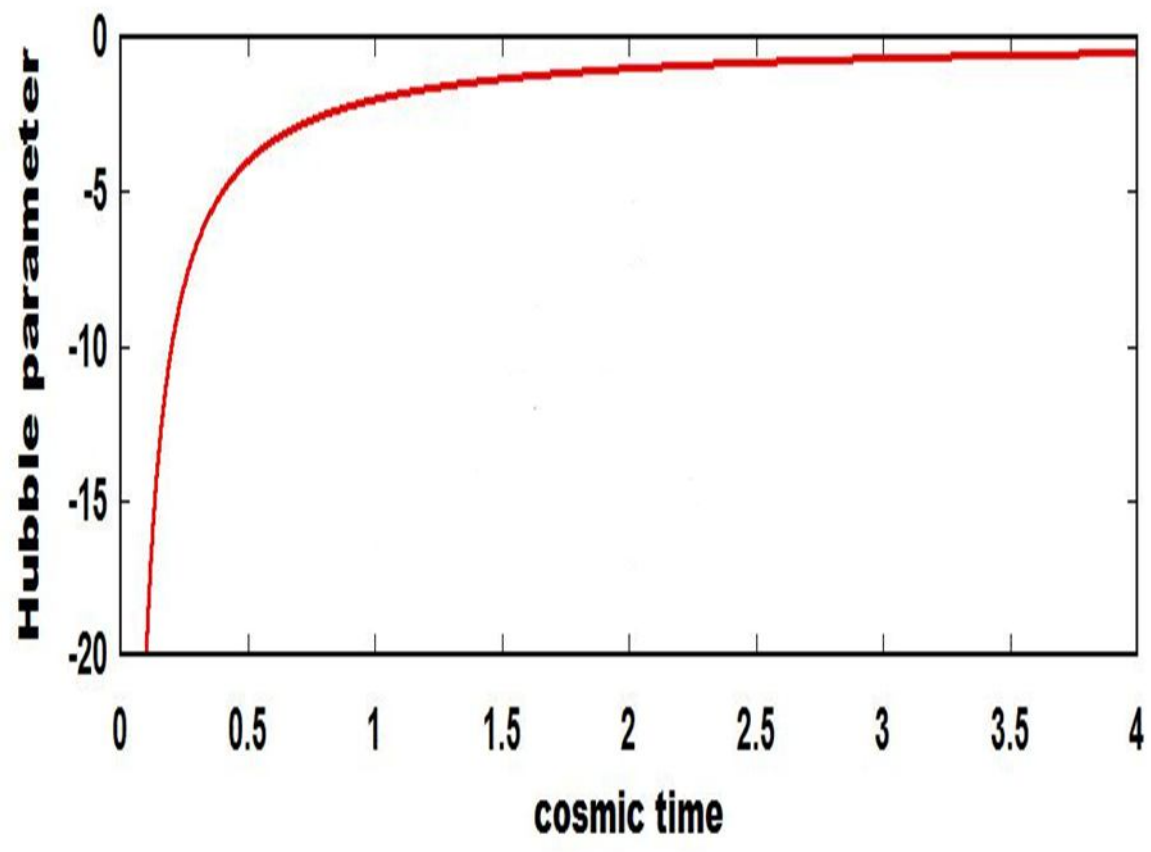

**Fig. 3 (b) Variation of Hubble parameter vs. cosmic time for $\omega = 1$.**

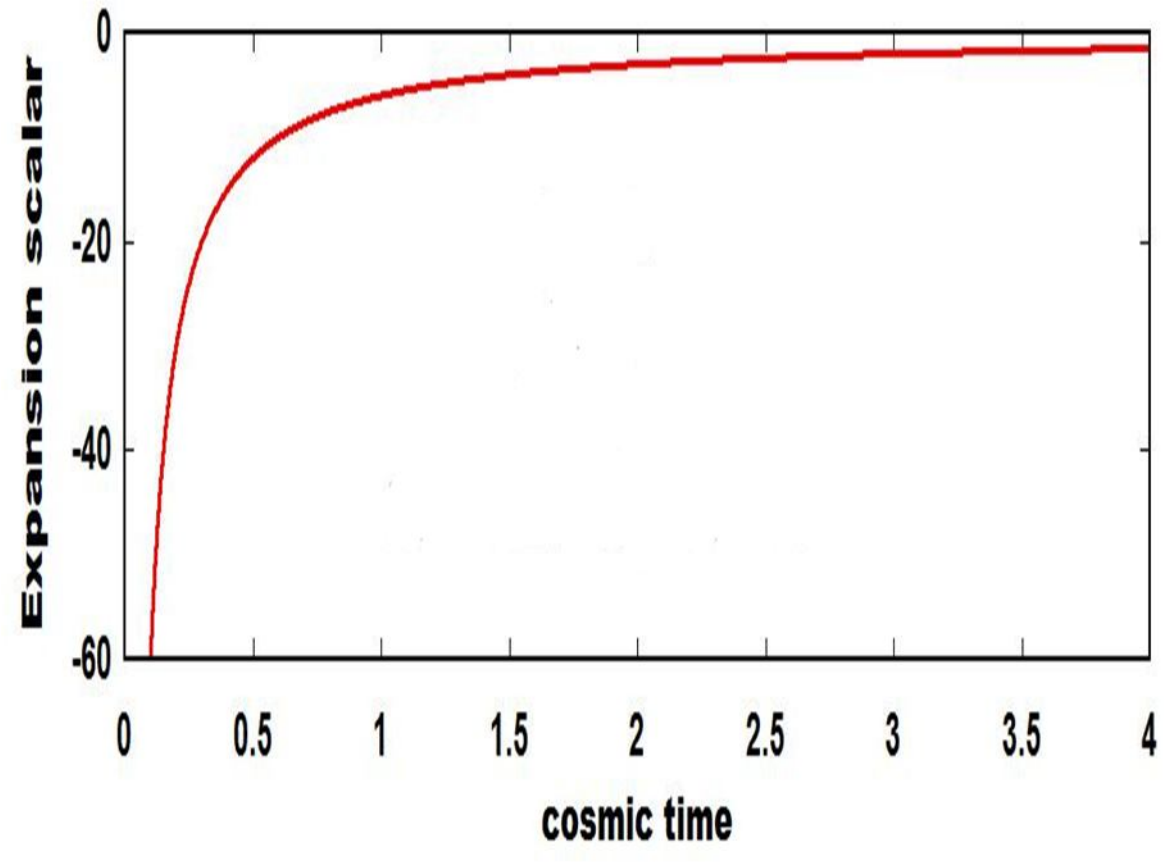

**Fig. 3 (c) Variation of expansion scalar vs. cosmic time.**

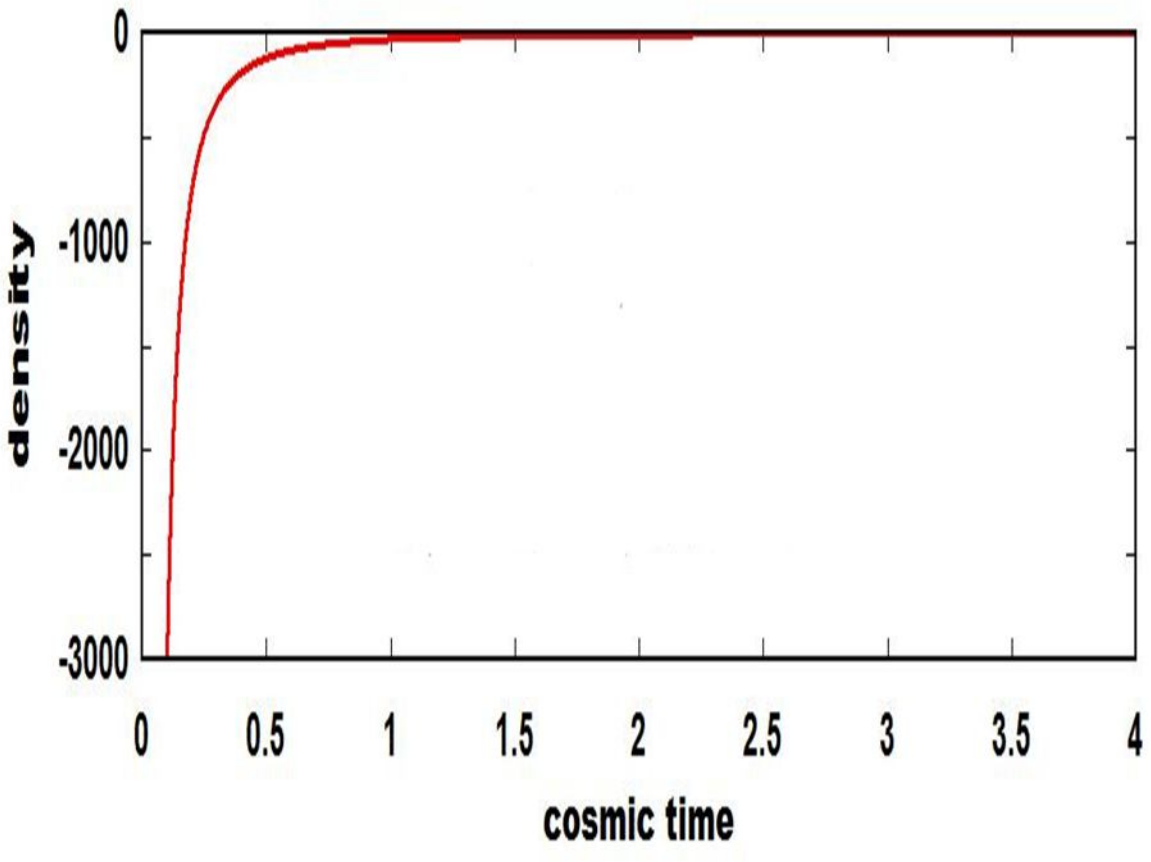

**Fig. 3 (d) Variation of density vs. cosmic time.**

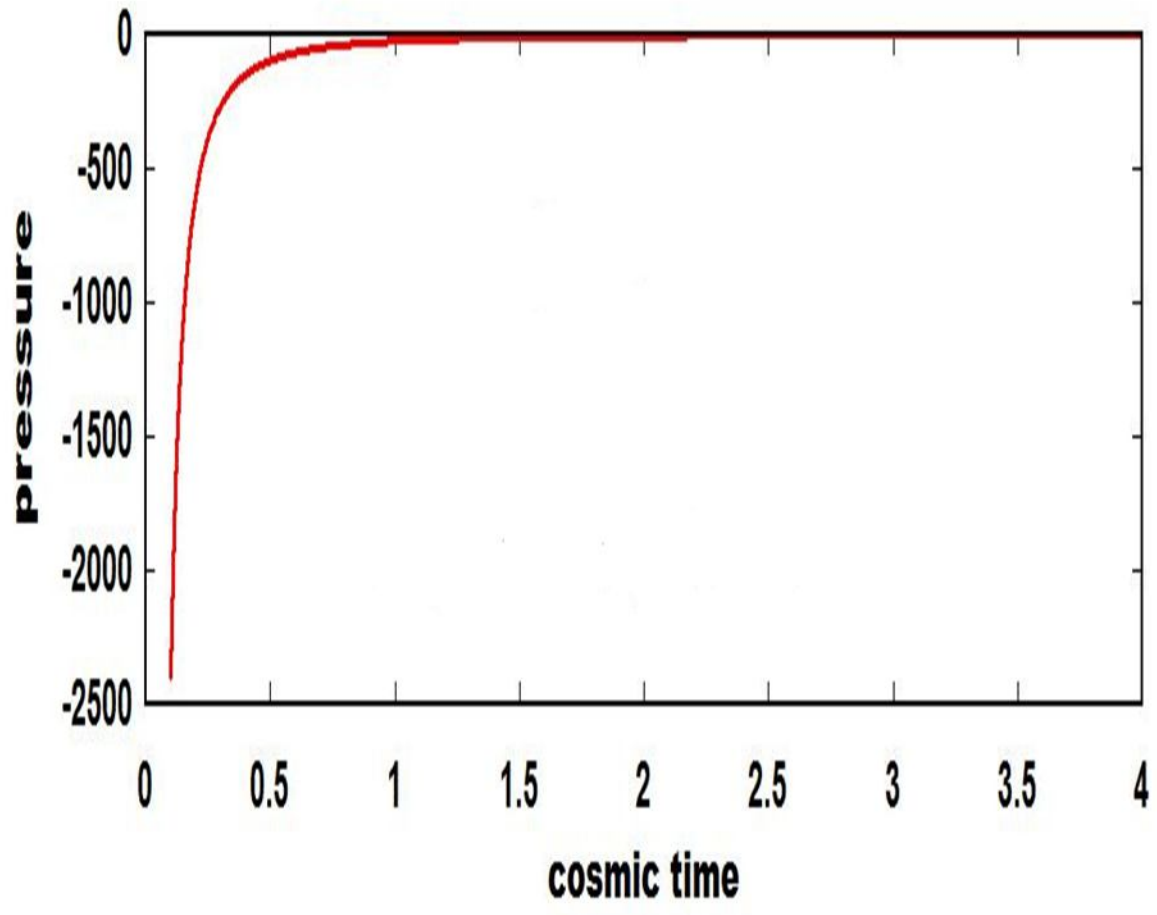

Fig. 3 (e) Variation of pressure vs. cosmic time for $\omega = 1$.

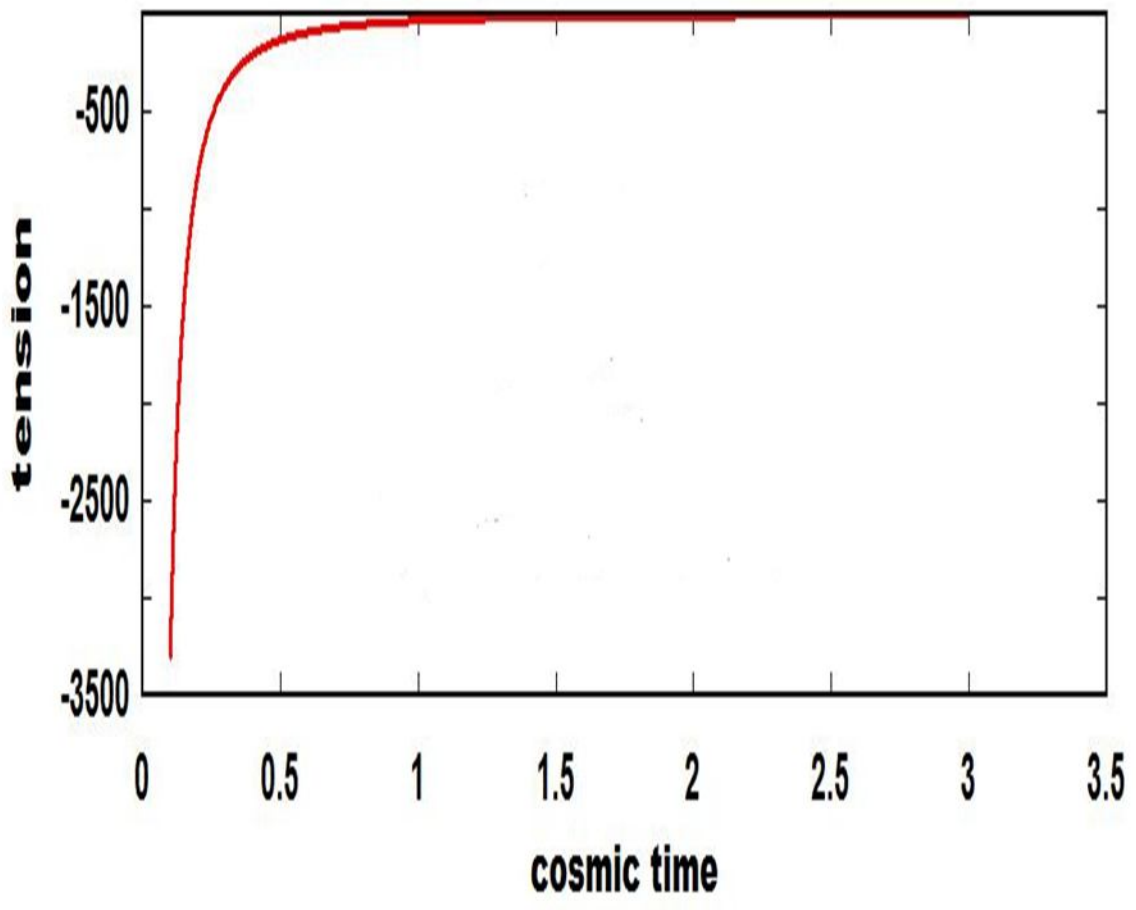

Fig .2 (f) Variation of tension vs. cosmic time for $\omega = 1$.

At the origin of the universe, volume of the model is undefined. It goes on decreasing as time increases and approaches to zero for large values of time as shown in the Fig. 3(a), In this case we obtained contracting universe. From equation (71), the deceleration parameter is negative throughout the evolution of the universe, which results into the accelerating universe. The Hubble parameter and expansion scalar are negative valued increasing function of cosmic time and vanishes at large time as shown in Fig. 3 (b) and Fig. 3(c). From Fig. 3 (d) and Fig. 3 (e), It is observed that initially when time is zero then density and pressure both approach to negative infinity. Both increase with time and vanish for large values of time. The energy density increases due to fractal geometry causes dissipation. Fig. 3(f) shows that the tension of the domain walls approaches to negative infinity at $t = 0$. The tension of domain walls increases with time and approaches to zero for large values of time, hence domain walls are invisible in this model and will be vanished in large time. The model violates the null energy condition; hence the present model is not in agreement with present universe.

## 5. Conclusion

In the present paper, we have studied the flat Friedmann-Robertson –Walker model in the frame work of fractal cosmology. The domain walls are taken as source of matter. We have discussed three different cases depending on value of fractal parameter $(\omega)$ and cosmic time$(t)$. The contracting and accelerating universe is obtained in case-I and case-III. The null energy condition is violated in both the cases; these models are not compatible with present universe.

In case-II, we have obtained the Expanding accelerated universe. At $t = 0$ the expansion of universe starts with a Big -Bang singularity. These observations are well suited with the realistic model. It is found that, the domain walls could exist in the early epoch and will be vanished in the far future. The obtained models are free from the gravitational collapse. These result matches with the results obtained by Katore et al. [38]. The behavior of domains walls is in according to the Zeldovich [49]. The existence of domain walls in the initial epoch contradicts to the results obtained by Adhav et al. [44], Reddy and Naidu [48]. We have performed the analysis under no cosmological constant and in UV regime.

**Acknowledgements**

The authors are would like to thank the anonymous referee for his/her valuable suggestions that have significantly improved the quality of the paper ae well as its presentation.

## REFERECES

1. Oyvind G., Sigbjorn H. (2007), New York, Springer
2. Riess R.G., Filippenko A.V., Challis P., Clocchiattia A., Diercks A. et al. (1998), The Astronomical Journal 116(3): 1009-1038
3. Riess R.G., Nugent P.E., Gillil R. L., Schmidt B.P., Torny J. et al. (2001), The Astrophysical Journal 560(1): 49-71
4. Riess R.G., Strolger L.G., Torny J., Casertano S., Ferguson H.C. et al. (2004), The Astronomical Journal 607(2): 665-687
5. Riess R.G., Strolger L.G., Casertano S., erguson H. C. et al. (2007), The Astronomical Journal 659(1): 98-121
6. Perlmutter S., Aldering G., Goldhaber G., Knop R.A., Nugent P.E. et al (1999)., The Astrophysical Journal 517(2): 5659-586
7. Scott F. Daniel, Robert R. Caldwell, Asantha Cooray, and Alessandro Melchiorri (2008), Phys. Rev. 77(10), arXiv:0802.1068 v1 [astro-ph]
8. Bennett C.L. et al. (2003), Astro Phys. J. Suppl.148,1
9. Max Tegmark et al. (2004), Phys. Rev.69(10): 103501
10. Brans C., Dicke R.H. (1961) Physical Review Journal Archive 124(3): 925-935
11. Barber G.A. (1982), Gen. Rel. Grav. 14,117
12. Saez and Ballester (1986), Phys. Lett. A 113,467
13. Buchdahl H.A. (1970), Mon. Not. R. astr. Soc 158: 1-8
14. Tiberiu Harko, Francisco SN Lobo et al. (2011), Physical Review D 84(2): 024020
15. Samanta G. S., R-Myrzakulov (2016), arXiv: 1611.02935 v2 [gr-qc].
16. Poonam Agrawal, Pawar D. D. (2018), Indian Journal of Science and Technology, Transactions A: Science 42(3):1639-1645
17. Sezgin Aygun, Pradyumn Kumar Sahoo, Binaya K. Bishi (2018), Gravitation and Cosmology 24(3): 302-307
18. I.Yilmaz and C. Atkas (2007) Chin. J. Astron. Astrophys 7,757
19. Pawar D. D., Bayaskar S. N., Patil V. R. (2009), Bulg. J. Phys. 36: 68-75
20. Katore S. D., Hatkar S. P. (2016), Prog. Exp. Phys. 2016(3):033E01
21. Agrawal P. K., Pawar D. D. (2017), New Astronomy 54: 56-60
22. Pradhan A., Rai V., Otarod S. (2006), Fizika B 15(57)
23. Patil V. R., Pawar D. D., Deshmukh A. G. (2010) Romanian Reports in Physics 62(4): 722-730

24. Ambjrn J., Jurkiewicz J., Loll R. (2005), Phys. Rev.Lett.95, 171301, arXiv: hep-th/0505113
25. Lauscher O., Reuter M. (2005), JHEP 10,050, arXiv: hep-th/0508202v1
26. Modesto L. (2009), class Quantum Gravity 26, 24002. arXiv: 0812.2214v1[gr-ac]
27. Linde A. D. (1986), Physics Lett. B.175(4): 395-400
28. Nottale Laurent (1992) Intl J Mod Phys A, 7(20): 4899-936
29. Calcagni G. (2010) Journal of High Energy Physics 03,120
30. Karami K., Mubasher Jamil, Ghaffari S., Fahimi K. (2012), Canadian journal of physics 91(10): 770-776
31. Mustafa Salti, Murat Korunur, Irfan Acikgoz (2014), European Physics Journal Plus 129(95): 1-13
32. Dipanjana Das, Sourav Dutta, Abdulla Al Mamon, Subenoy Chakraborty (2018), The European Physical Journal C 78: 849
33. Lemets O. A., Yerokhin D.A. (2012), arXiv:1202.3457v3[astro-ph.co]
34. Hosseienkhan H.i, Youesefi H., Azimi N. (2018), International Journal of Geometric Methods in Modern Physics. 15(12), (1850200)
35. Joyce M., Anderson P.W., Montuori M., Pietronero L.and Sylos Labini (2000), Europhysics Letters 15(3).
36. Naidu R. L., K. Dasu Naidu, T. Ramprasad and D.R.K Reddy (2013), Global Journal of Science Frontier Research Physics and Space Science. 13(3)
37. Joachim Schroter (2012), Adv. Theor. Math. Phys. 16: 393-419
38. Katore S. D., Hatkar S. P., Baxi R. J. (2016), Chines Journal of Physics. 54: 563-573
39. Boyanovsky D., Brahm D. E., A. Gonzalez-Rulz, Holman R. and Takakurva F. I. (1995), arXiv:hep-ph/9501380v3
40. Rosales J.J. and Tkach V.I. (2010), Phys. Rev. D 82: 107502
41. Vilenkin A. (1983), Phys. Lett. B 133: 177
42. Ipser J. and Sikivie P (1984), Physical Review D. 30 (4): 712-719
43. Rahaman F., Kalam M., Mondal R. (2006), Astrophys. Space Sci. 305: 337-340
44. Adhav K. S., Nimkar A. S., Ghate H. R, Pund A. M. (2008), Rom. Journ. Phys. 53(7-8): 909-915
45. Sahoo P. K. and Bivudutta Mishra (2013), Journal of Theoretical and Applied Physics. 7:12
46. Shaikh A. Y., Wankhade K. S. (2018), Int. J. S. Res. Sci. Technol. 4(10): 134-141
47. Wang Anzhong (1994), Modern Physics Letters A. 9(39): 3605-3609
48. Reddy D.R.K, Naidu R. L. (2007), Int. J. Theor. Phys. 46: 2788-2794.
49. Zeldovich Y. B, Kobzarev I. Y., Okun L. B. (1975), Sov. Phys. JETP. 40(1):1-5